\documentclass[useAMS,usegraphicx]{mn2e}

\title[AGN and starburst in ULIRGs]{Spectral decomposition of starbursts
and AGNs in 5--8 $\bmu$m \textit{Spitzer} IRS spectra of local ULIRGs}
\author[E. Nardini et al.]{E.~Nardini,$^1$ G.~Risaliti,$^{2,3}$ M.~Salvati,$^2$ E.~Sani,$^1$
\newauthor M.~Imanishi,$^4$ A.~Marconi$^1$ and R.~Maiolino$^5$\\
$^1$ Dipartimento di Astronomia, Universit\`a di Firenze, L.go E. Fermi 2, 50125 Firenze, Italy.
{E-mail: nardini@arcetri.astro.it}\\
$^2$ INAF - Osservatorio Astrofisico di Arcetri, L.go E. Fermi 5, 50125 Firenze, Italy\\
$^3$ Harvard-Smithsonian Center for Astrophysics, 60 Garden St. Cambridge, MA 02138 USA\\
$^4$ National Astronomical Observatory, 2-21-1, Osawa, Mitaka, Tokyo 181-8588, Japan\\
$^5$ INAF - Osservatorio Astronomico di Roma, via di Frascati 33, 00040 Monte Porzio Catone (RM), Italy}

\begin{document}

\date{Released Xxxx Xxxxx XX}

\pagerange{\pageref{firstpage}--\pageref{lastpage}} \pubyear{2002}

\maketitle

\label{firstpage}

\begin{abstract}
We present an analysis of the 5--8~$\mu$m \textit{Spitzer}-IRS
spectra of a sample of 68 local Ultraluminous Infrared Galaxies (ULIRGs).
Our diagnostic technique allows a clear separation of the
active galactic nucleus (AGN) and starburst (SB) components
in the observed mid-IR emission, and a simple analytic model provides
a \textit{quantitative} estimate of the AGN/starburst contribution
to the bolometric luminosity. We show that AGNs are $\sim30$ times
brighter at 6~$\mu$m than starbursts with the same bolometric luminosity,
so that even faint AGNs can be detected.
Star formation events are confirmed as the dominant power source
for extreme infrared activity, since $\sim85\%$ of ULIRG luminosity
arises from the SB component. Nonetheless an AGN is
present in the majority (46/68) of our sources.


\end{abstract}


\section{Introduction}

Ultraluminous Infrared Galaxies (ULIRGs, $L_\mathit{IR}>10^{12} L_\odot$)
are the local counterparts of the high-redshift
objects dominating the cosmic background in the far-infrared
and millimetric bands. Unveiling the nature of their energy
source is fundamental in order to understand the
star formation history and the obscured AGN activity in the
distant Universe.

Since their discovery, several infrared
indicators have been proposed to determine whether the central
engine in ULIRGs is an AGN or a starburst (SB). The presence of
high-ionization lines in the mid-IR spectra of ULIRGs points
to AGN activity, while intense PAH emission features are typical of
starburst environments (Genzel et al. 1998; Laurent et al. 2000).
Recently, the absorption feature of amorphous silicate
grains centered at 9.7~$\mu$m has also been used together
with the PAH emission to assess the nature of
the obscured power source (Spoon et al. 2007).
An alternate way to disentangle AGNs and SBs in ULIRGs
has been proposed by Risaliti et al. (2006, hereafter R06),
based on the separation of the two continuum components in
3--4~$\mu$m spectra. This method has been successfully applied
to a sample of $\sim$50 nearby ULIRGs (Risaliti, Imanishi \&
Sani 2007, \textit{submitted}) and provided an estimate of the
average AGN/SB contribution to ULIRGs.
The key reason for using the continuum emission at
$\lambda \simeq$3--4~$\mu$m as a diagnostic is the
difference between the 3-$\mu$m to bolometric ratios in
AGNs and starbursts ($\sim$two orders of magnitude larger in
the former). This makes the detection of the AGN component possible
even when the AGN is heavily obscured and/or bolometrically weak compared
to the starburst. However the original prescription is limited
by the low quality of the available L-band spectra
of ULIRGs (R06, Imanishi et al. 2006),
which makes the results on individual sources highly
uncertain, except for the $\sim$10--15 brightest objects.

At present we have extended the analysis to the 5--8~$\mu$m
spectral band, using the observations of the IRS instrument
(Houck et al.~2004) onboard \textit{Spitzer}.
We disentangled the AGN and SB
contributions to the observed 5--8~$\mu$m emission of
ULIRGs by combining average spectral templates representing
the different properties of the two physical processes at work.
The high quality of \textit{Spitzer}-IRS data, in addition to
the relatively low dispersion of the intrinsic continuum
properties of both AGNs and starbursts in this spectral range,
allows a much more accurate determination of the AGN/SB components
than possible at other wavelengths (e.g. X-rays) or with other
diagnostic methods based on emission lines.
In this paper we present our decomposition method,
and discuss a simple analytical model providing
a \textit{quantitative} estimate of the AGN/SB contribution
to the bolometric luminosity of each source.

\section{Observations and Data Reduction}

In order to perform a detailed study of a representative sample of ULIRGs in
the local Universe, we selected 68 sources with $z<0.15$ and
a 60-$\mu$m flux density $f_{60}>1$~Jy.
Most of the objects are
taken from the IRAS ULIRG 1~Jy sample (Kim \& Sanders 1998),
but a few more sources in the southern {hemisphere}
have also been included. The flux limit at 60~$\mu$m 
ensures an unbiased selection with respect to the relative AGN/SB 
contributions.

IRS observations were obtained within three different
programs: PID 105 (PI J.R.Houck), PID 2306
(PI M.Imanishi), PID 3187 (PI S.Veilleux).
The coadded images provided by the \textit{Spitzer Science
Center} (after the treatment with pipeline version S13.0) have been
background-subtracted by differencing the two observations
in the nodding cycle. The spectra have been extracted and calibrated
following the standard procedure for point-like sources with the
package \texttt{SPICE}. The flux uncertainties have been estimated from source and
background counts (in $e^-/s$). Finally, we performed a smooth
connection between the Short-Low spectral orders, with no necessity
of relative scaling.

Out of the 68 spectra, we already published 48 in Imanishi et al.~2007.
Six more spectra are shown in Armus et al.~2006, 2007. The remaining 14 spectra are analyzed
here for the first time and will be fully presented in a forthcoming paper
(Nardini et al. 2008, \textit{in prep.}).

\section{The 5--8~$\bmu$\lowercase{m} AGN/SB separation}
\label{3}
Despite the diversity of the global IRS spectra of pure AGNs and pure SBs,
and the complexity of the physics involved, little
dispersion is seen at wavelengths shortward of the
9.7$~\mu$m silicate feature. This makes possible the use
of universal AGN/SB templates to reproduce the spectral
properties of ULIRGs in the 5--8~$\mu$m interval.
In the following we describe the templates adopted in our model.

\begin{figure}
\includegraphics[width=8.5cm]{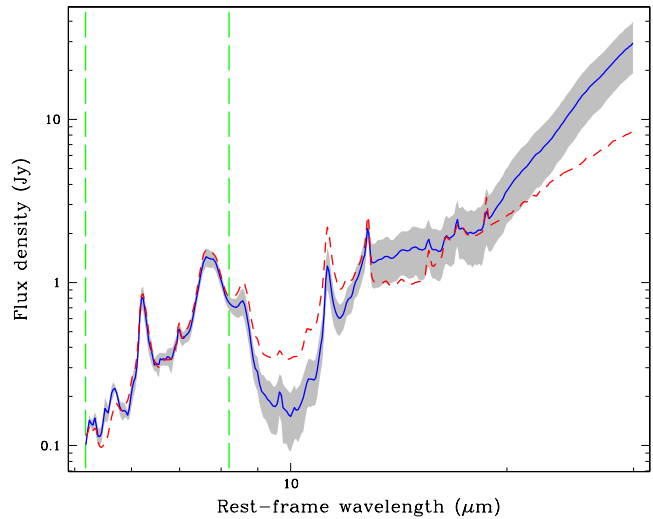}
\caption{Comparison between the SB template of B06
(\textit{dashed red line}) and our template (\textit{solid blue line}),
constructed from the emission of the 5 brightest SB-dominated ULIRGs. The \textit{shaded
area} shows the \textit{1-$\sigma$ rms dispersion} in the 5 ULIRG spectra. The vertical \textit{long-dashed green lines}
enclose the fitting region.}
\label{f1}
\end{figure}

\textit{Starburst.} The mid-IR spectral features of local SB
galaxies show very little variations from one object to another
in the 5--8~$\mu$m wavelength range, while larger
differences are present in the $\sim$9--30~$\mu$m band
(Brandl et al. 2006, hereafter B06).
In order to check if this is the case in the ULIRG luminosity range
as well, we analyzed the sources in our sample estabilished to be
starburst-dominated by multi-band studies.
We did not find significant variations among the spectra,
concluding that a fixed template can be used to represent
the 5--8~$\mu$m SB component in local ULIRGs. We built such
template using the five brighest objects among the pure SBs in
our sample (IRAS~10190$+$1322, IRAS~12112$+$0305, IRAS~17208$-$0014,
IRAS~20414$-$1651 and IRAS~22491$-$1808), whose underlying continuum
has been reproduced by a power law and normalized at 6~$\mu$m before adding.
Our SB template is shown in Fig.\ref{f1}, together with its dispersion
in the whole IRS spectral band and the B06 template. The little spectral
dispersion below 8~$\mu$m among SBs of different luminosity
(to be compared with their large differences at longer wavelengths)
is in itself an interesting result, which should be fully investigated through
detailed emission and radiative transfer models. Concerning this we only
notice that such a remarkable similarity can result from the spatial
integration over a large number of individual star-forming regions.

\textit{AGN.} Our recent L-band analysis of bright ULIRGs
shows that the intrinsic AGN emission is due to hot dust
grains and the flux density is well described by a
featureless power law with a fixed spectral index:
$f_\nu \propto \lambda^{1.5}$. Here we adopt the same
spectral shape up to 8~$\mu$m, in agreement with new
\textit{Spitzer} observations of a large sample of local
type~1 quasars (Netzer et al. 2007).

An active nucleus is much more compact than a circumnuclear
starburst region. As a consequence, the near-IR
radiation due to thin dust reprocessing can be itself strongly
reddened because of a compact absorber along the line of sight.
We therefore introduce an exponential attenuation factor
$e^{-\tau(\lambda)}$, where the optical depth follows the
conventional law $\tau(\lambda) \propto \lambda^{-1.75}$ (Draine 1989).
A similar correction is not needed in the SB template.
We stress that this does not imply that the starburst spectrum
is not affected by inner obscuration; the possible effects of this obscuration
(which are clear at longer wavelengths, e.g. in the silicate absorption features
at 9.7 and $\sim$18~$\mu$m) are however
already accounted for in the adopted observational template.

Summarizing, the different contributions to the observed energy output
of a ULIRG can be parametrized as follows:
\begin{equation}
f_\nu^\mathit{obs}(\lambda)=f_6^\mathit{int}\left[(1-\alpha_6)u_\nu^\mathit{sb}+\alpha_6 u_\nu^\mathit{agn} e^{-\tau(\lambda)}\right]
\end{equation}
where $\alpha_6$ is the AGN contribution to the
5--8~$\mu$m intrinsic flux density $f_6^\mathit{int}$, while $u_\nu^\mathit{sb}$ and
$u_\nu^\mathit{agn}$ are the SB and AGN templates normalized at 6~$\mu$m.
Apart from the flux normalization, our model contains only two free
parameters, i.e. $\alpha_6$ and the optical depth to the AGN $\tau$(6~$\mu$m).
They are both shown in Tab.\ref{t1}.

Additional high-ionization emission lines and molecular absorption features
(due to ices and aliphatic hydrocarbons), whenever
present, were fitted by means of gaussian profiles
except for the water ice absorption at $\sim$6~$\mu$m,
reproduced with the laboratory profile from the Leiden
database corresponding to pure H$_2$O ice at 30~K.

\begin{figure}
\includegraphics[width=8.5cm]{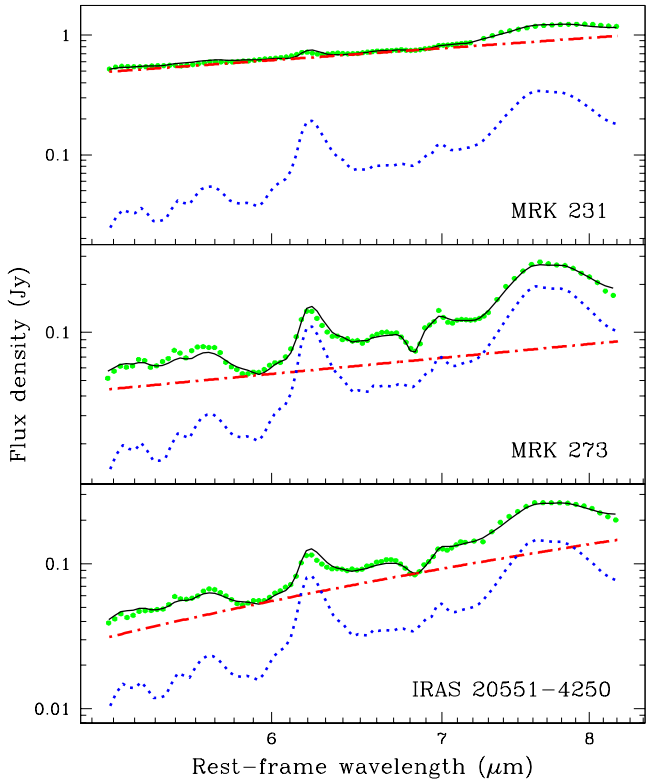}
\caption{Three representative examples of the
typical 5--8~$\mu$m spectral shapes of ULIRGs.
The differences among the spectra are entirely due to the different AGN
contribution and its obscuration.
Whenever
the AGN is the dominant power source, as in Mrk~231, a
strong continuum almost obliterates the PAH features.
On the contrary, in Mrk~273 the AGN is fainter and the
spectral outline of a starburst is clearly identified.
A similar spectrum is exhibited by IRAS~20551$-$4250,
but the features are less prominent and the continuum
is steeper: this source harbours an obscured AGN.
In each panel, in addition to the data
(\textit{green filled circles}) and their best fits
(\textit{black thin line}), we have included the
reddened AGN (\textit{red dot-dashed line}) and starburst
(\textit{blue dotted line}) components.}
\label{f2}
\end{figure}

Fig.\ref{f2} shows the spectral decomposition of three
representative ULIRGs. 
In spite of the great diversity of the observed
spectra, our simple model provides a good fit of each spectrum
in the sample: the residuals from the best fits are in all sources
smaller than 10\% at all wavelengths (though the fits are not
formally acceptable in a statistical sense, with a reduced $\chi^2\ga$2, due
to the small error bars and the remaining unfitted minor features).
In particular, both the PAH emission and the continuum
are always well reproduced: this implies that the large variations
in the 5--8~$\mu$m spectral shape of ULIRGs are entirely due
to the AGN contribution and its obscuration.
A detailed analysis of the results for each source
and a physical interpretation of peculiar cases are
the subject of a forthcoming paper (Nardini et al. 2008,
\textit{in prep.}).


\begin{table*}
\begin{center}
\caption{Spectral parameters for the 68 sources in our sample.
$\alpha_6$: AGN contribution to the intrinsic
continuum emission at 6~$\mu$m (in percent). $\tau$:
Optical depth of the AGN component at 6~$\mu$m (we assume
$\tau=0$ for the sources with no detected AGN).
$\alpha_\mathit{bol}$: AGN contribution to the bolometric luminosity
(in percent). The errors in $\alpha_\mathit{bol}$ are due to the
statistical uncertainty both in the flux amplitude of the AGN/SB
components and in the ratios $R^\mathit{agn}$ and $R^\mathit{sb}$. The systematic
effects are discussed in the text. $O/X/L$:
SB/AGN/LINER classification based on optical, X-ray and L-band spectroscopy. A: AGN, L: LINER,
A*: AGN, tentative detection.
References:
$^1$:~Veilleux et al.~1999, $^2$:~Veilleux et al.~1995, $^3$:~Duc et al.~1997, {$^4$:~Iwasawa et al.~2005, $^5$:~Severgnini et al.~2001, $^6$:~Franceschini et al.~2003, $^7$:~Balestra et al.~2005, $^8$:~Vignati et al.~1999, $^9$:~Imanishi et al.~2003, $^{10}$:~Imanishi et al.~2006, $^{11}$:~Risaliti et al.~2006, $^{12}$:~Sani et al.~2007}. $^\dagger$: Sources with simultaneous
$\alpha_\mathit{bol}>0.2$ and $\tau>1$.}
\label{t1}
\begin{scriptsize}
\begin{tabular}{lcccc|lcccc}
\hline
Source            &$\alpha_6$& $\tau$        & $\alpha_\mathit{bol}$   & $O/X/L$ & Source            &$\alpha_6$& $\tau$ & $\alpha_\mathit{bol}$ & $O/X/L$ \\
\hline
ARP 220           & 75$\pm$1 & 1.40$\pm$0.01 & $9.8^{+3.7}_{-2.7}$ &L$^1$/SB$^4$/SB$^{10}$ & IRAS 14197$+$0813 & 75$\pm$2 & 2.10$\pm$0.14 & $10^{+5}_{-4}$      &--/--/--         \\
IRAS 00091$-$0738$^\dagger$ & 90$\pm$1 & 1.81$\pm$0.04 & $25^{+8}_{-7}$      &SB$^1$/--/--         & IRAS 14252$-$1550 & $<20$ & $<0.04$          & $<0.9$              &L$^1$/--/SB$^{10}$  \\
IRAS 00188$-$0856 & 96$\pm$1 & 0.60$\pm$0.02 & 45$\pm$9            &L$^1$/--/A*$^{10}$    & IRAS 14348$-$1447 & $51^{+3}_{-5}$ & $<0.09$ & $3.7^{+2.1}_{-1.5}$ &L$^1$/--/SB$^{11}$  \\
IRAS 00456$-$2904 & $<1.0$   & 0             & $<0.04$             &SB$^1$/--/--         & IRAS 15130$-$1958 & $91^{+1}_{-3}$ & $<0.01$ & $28^{+9}_{-11}$     &A$^1$/--/A$^{10}$  \\
IRAS 00482$-$2721 & $<54$    & $<0.04$       & $<4.1$              &L$^1$/--/--          & IRAS 15206$+$3342 & 52$\pm$1 & 0.30$\pm$0.02 & $3.9^{+1.7}_{-1.2}$ &SB$^1$/--/SB$^{10}$  \\
IRAS 01003$-$2238$^\dagger$ & 96$\pm$1 & 1.58$\pm$0.02 & 49$\pm$9            &SB$^1$/--/--         & IRAS 15225$+$2350 & 89$\pm$1 & 0.77$\pm$0.02 & $22^{+7}_{-6}$      &SB$^1$/--/SB$^{10}$  \\
IRAS 01166$-$0844 & 87$\pm$1 & 1.19$\pm$0.06 & $20^{+7}_{-6}$      &SB$^1$/--/--         & IRAS 15250$+$3609 & 94$\pm$1 & 0.91$\pm$0.01 & $35^{+9}_{-8}$      &L$^2$/SB$^6$/--  \\
IRAS 01298$-$0744$^\dagger$ & 98$\pm$1 & 1.79$\pm$0.02 & $74^{+8}_{-9}$      &SB$^1$/--/--         & IRAS 15462$-$0450 & $90^{+1}_{-16}$ &$<0.01$ & $25^{+10}_{-18}$    &A$^1$/--/--  \\
IRAS 01569$-$2939 & 85$\pm$1 & 1.13$\pm$0.04 & $17^{+6}_{-5}$      &SB$^1$/--/--         & IRAS 16090$-$0139 & 89$\pm$1 & 0.69$\pm$0.01 & $23^{+7}_{-6}$      &L$^1$/--/A*$^{10}$  \\
IRAS 02411$+$0353 & $<17$    & $<0.01$       & $<0.8$              &SB$^1$/--/--         & IRAS 16156$+$0146 & 90$\pm$1 & 0.37$\pm$0.01 & $25^{+8}_{-6}$      &A$^1$/--/--  \\
IRAS 03250$+$1606 & $<3.4$   & $<0.11$       & $<0.2$              &L$^1$/--/A*$^{10}$    & IRAS 16468$+$5200 & 85$\pm$1 & 0.77$\pm$0.02 & $18^{+6}_{-5}$      &L$^1$/--/SB$^{10}$  \\
IRAS 04103$-$2838 & 56$\pm1$ & 0.08$\pm$0.02 & $4.5^{+2.0}_{-1.4}$ &L$^1$/--/--          & IRAS 16474$+$3430 & $<4.9$ & $<0.01$         & $<0.2$              &SB$^1$/--/A*$^{10}$  \\
IRAS 05189$-$2524 & $91^{+1}_{-4}$ & $<0.01$ & $28^{+8}_{-12}$     &A$^1$/A$^5$/A$^{12}$ & IRAS 16487$+$5447 & 21$\pm$1 & $<0.04$ & $1.0^{+0.5}_{-0.4}$ &L$^1$/--/A*$^{10}$  \\
IRAS 08572$+$3915 & 99$\pm$1 & 0.44$\pm$0.01 & $85^{+5}_{-6}$      &L$^1$/--/A$^{10}$     & IRAS 17028$+$5817 & $<1.2$ & 0               & $<0.05$             &L$^1$/--/A*$^{10}$  \\
IRAS 09039$+$0503 & 61$\pm$1 & 0.71$\pm$0.04 & $5.5^{+2.5}_{-1.7}$ &L$^1$/--/A*$^{10}$    & IRAS 17044$+$6720 & 91$\pm$1 & 0.32$\pm$0.01 & $26^{+8}_{-6}$      &L$^1$/--/A$^{10}$  \\
IRAS 09116$+$0334 & $<1.8$   & $<0.01$       & $<0.07$             &L$^1$/--/A*$^{10}$    & IRAS 17179$+$5444 & 84$\pm$1 & 0.31$\pm$0.02 & 16$^{+6}_{-4}$      &A$^1$/--/A$^{10}$  \\
IRAS 09539$+$0857$^\dagger$ & 91$\pm$1 & 1.85$\pm$0.03 & $27^{+8}_{-7}$      &L$^1$/--/SB$^{10}$      & IRAS 17208$-$0014 & $<7.9$ & $<0.01$         & $<0.4$              &L$^3$/SB$^6$/SB$^{11}$  \\
IRAS 10190$+$1322 & $<0.3$   & 0             & $<0.02$             &SB$^1$/--/SB$^{10}$     & IRAS 19254$-$7245 & 89$\pm$1 & 0.21$\pm$0.08 & $23^{+9}_{-7}$      &A$^3$/A$^6$/A$^{11}$  \\
IRAS 10378$+$1109 & $73^{+1}_{-9}$ & $<0.01$ & $9.0^{+3.7}_{-4.6}$ &L$^1$/--/A*$^{10}$    & IRAS 20100$-$4156 & 86$\pm$1 & 0.47$\pm$0.02 & $19^{+6}_{-5}$      &SB$^3$/A$^6$/SB$^{11}$  \\
IRAS 10485$-$1447 & 60$\pm$1 & 0.16$\pm$0.03 & $5.3^{+2.4}_{-1.7}$ &L$^1$/--/A*$^{10}$    & IRAS 20414$-$1651 & $<2.2$ & $<0.07$         & $<0.09$             &SB$^1$/--/SB$^{10}$  \\
IRAS 10494$+$4424 & $<0.4$   & 0             & $<0.02$             &L$^1$/--/A*$^{10}$    & IRAS 20551$-$4250$^\dagger$ & 90$\pm$1 & 1.19$\pm$0.01 & $25^{+8}_{-6}$      &L$^3$/A$^6$/A$^{12}$  \\
IRAS 11095$-$0238$^\dagger$ & 94$\pm$1 & 1.22$\pm$0.01 & $35^{+9}_{-8}$      &L$^1$/--/SB$^{10}$      & IRAS 21208$-$0519 & $<0.9$ & 0               & $<0.04$             &SB$^1$/--/SB$^{10}$  \\
IRAS 11130$-$2659 & 84$\pm$1 & 1.15$\pm$0.02 & $16^{+6}_{-5}$      &L$^1$/--/--          & IRAS 21219$-$1757 & $99^{+1}_{-4}$ & $<0.01$ & $83^{+13}_{-48}$    &A$^1$/--/A$^{10}$  \\
IRAS 11387$+$4116 & $<0.8$   & 0             & $<0.03$             &SB$^1$/--/SB$^{10}$     & IRAS 21329$-$2346 & 43$\pm2$ & $<0.07$        & $2.7^{+1.3}_{-0.9}$ &L$^1$/--/A*$^{10}$  \\
IRAS 11506$+$1331 & $54^{+1}_{-21}$ &$<0.01$ & $4.2^{+1.9}_{-2.9}$ &SB$^1$/--/A*$^{10}$   & IRAS 22206$-$2715 & $<9.3$ & $<0.17$         & $<0.4$              &SB$^1$/--/--  \\
IRAS 12072$-$0444$^\dagger$ & 94$\pm$1 & 1.06$\pm$0.01 & $37^{+9}_{-8}$      &A$^1$/--/A$^{10}$   & IRAS 22491$-$1808 & $<1.4$ & 0               & $<0.06$             &SB$^1$/SB$^6$/--  \\
IRAS 12112$+$0305 & $<22$ & $<0.02$          & $<1.0$              &L$^1$/SB$^6$/SB$^{11}$  & IRAS 23128$-$5919 & 48$\pm$1 & 0.36$\pm$0.03 & $3.4^{+1.5}_{-1.0}$ &L$^3$/A$^6$/A$^{11}$  \\
IRAS 12127$-$1412 & 98$\pm$1 & 0.24$\pm$0.08 & $60^{+15}_{-14}$    &L$^1$/--/A$^{10}$     & IRAS 23234$+$0946 & $30^{+2}_{-3}$ & $<0.05$ & $1.6^{+0.8}_{-0.6}$ &L$^1$/--/SB$^{10}$  \\
IRAS 12359$-$0725 & $54^{+2}_{-43}$& $<0.01$ & $4.2^{+2.2}_{-3.9}$ &L$^1$/--/A*$^{10}$    & IRAS 23327$+$2913 & 73$\pm$1 & 0.86$\pm$0.03 & $9.1^{+3.8}_{-2.7}$ &L$^1$/--/SB$^{10}$  \\
IRAS 13335$-$2612 & $<0.6$   & 0             & $<0.02$             &L$^1$/--/--    & MRK 231           & 93$\pm$1 & $<0.12$       & $32^{+8}_{-7}$      &A$^1$/A$^6$/A$^{10}$  \\
IRAS 13454$-$2956 & $59^{+1}_{-28}$ &$<0.01$ & $5.0^{+2.2}_{-3.8}$ &A$^1$/--/--         & MRK 273           & $67^{+1}_{-4}$ & $<0.01$ & $7.0^{+2.8}_{-2.7}$ &A$^1$/A$^7$/A$^{10}$  \\
IRAS 13509$+$0442 & $<0.6$   & 0             & $<0.03$             &SB$^1$/--/SB$^{10}$     & NGC 6240     & $65^{+6}_{-8}$& 0.64$\pm$0.24 & $6.5^{+4.8}_{-3.1}$ &L$^2$/A$^8$/A$^{12}$  \\
IRAS 13539$+$2920 & $<0.4$   & 0             & $<0.02$             &SB$^1$/--/SB$^{10}$     & 4C +12.50         & 97$\pm$1 & 0.24$\pm$0.02 & $59^{+10}_{-11}$    &A$^1$/--/--  \\
IRAS 14060$+$2919 & $<0.4$   & 0             & $<0.02$             &SB$^1$/--/SB$^{10}$     & UGC 5101$^\dagger$          & 92$\pm$1 & 1.09$\pm$0.03 & $30^{+9}_{-8}$      &L$^2$/A$^9$/A$^{10}$  \\
\hline
\end{tabular}
\end{scriptsize}
\end{center}
\end{table*}

\section{AGN/SB bolometric contributions}
\label{4}

The large difference
between the 5--8~$\mu$m to bolometric ratios in AGNs and SBs
implies that this ratio is itself an indicator of AGN activity,
and can be used \textit{(a)} to test the consistency of our decomposition
method and \textit{(b)} to estimate the relative AGN and SB contributions
to the bolometric luminosity of our sample.
We define the 6-$\mu$m to bolometric ratio as:
\begin{equation}
R=\left(\frac{\nu_6 f_6^\mathit{int}}{F_\mathit{IR}}\right).
\label{e2}
\end{equation}
where $F_\mathit{IR}$ is the total infrared
flux, estimated as in Sanders \& Mirabel~(1996).
Since the integrated infrared luminosity
of ULIRGs coincides almost exactly with their bolometric
luminosity, $R$ is a fair approximation to the fraction of the total
energy output that is intrinsically emitted in the 5--8~$\mu$m range.
Reminding that the intrinsic AGN/SB contributions are $\alpha_6 f_6^\mathit{int}$ and
$(1-\alpha_6)f_6^\mathit{int}$ respectively, and decomposing $F_\mathit{IR}$ as $F_\mathit{IR}^\mathit{agn}+F_\mathit{IR}^\mathit{sb}$, a simple connection
between $R$ and $\alpha_6$ is brought out:
\begin{equation}
\label{e3}
R=\frac{R^\mathit{agn}R^\mathit{sb}}{\alpha_6 R^\mathit{sb}+(1-\alpha_6) R^\mathit{agn}},
\end{equation}
provided that $R^\mathit{agn}$ and $R^\mathit{sb}$, the equivalents of $R$ for
pure (unobscured) AGNs and pure SBs, are defined as in Eq.\ref{e2}.
We fitted the theoretical $R(\alpha_6)$ relation (Eq.\ref{e3})
to our data considering $R^\mathit{agn}$ and
$R^\mathit{sb}$ as free parameters, and found:
\begin{equation}
\label{e4}
\log R^\mathit{agn}=-0.49^{+0.12}_{-0.13} \hspace*{10pt} \textrm{and} \hspace*{10pt} \log R^\mathit{sb}=-1.93^{+0.03}_{-0.03}.
\end{equation}
We note that $R^\mathit{agn}$ turns out to be somewhat
higher than traditional estimates based on AGN spectral energy
distributions: for example, we derive $\log R^\mathit{agn} \sim -0.6$
from the SED of Elvis et al. (1994). This suggests that the local quasars
(mostly PG quasars) used to build the mentioned SED
can be contaminated to some extent by a starburst contribution, in
agreement with recent studies (Netzer et al. 2007).

According to Eq.\ref{e4}, AGNs are $\sim$30 times more luminous at 6~$\mu$m
than starbursts with the same bolometric luminosity.
We are now able to quantify the AGN contribution (${\alpha_\mathit{bol}=F_\mathit{IR}^\mathit{agn}/F_\mathit{IR}}$) to
the total infrared luminosity of each source:
\begin{equation}
\alpha_\mathit{bol}=\frac{\alpha_6}{\alpha_6+(R^\mathit{agn}/R^\mathit{sb})(1-\alpha_6)},
\label{e5}
\end{equation}
where $R^\mathit{agn}/R^\mathit{sb}\simeq28$.
The values of $\alpha_\mathit{bol}$ are listed in Tab.\ref{t1}.
Our estimates for the $\sim$15 brightest sources are in good agreement
with those of R06 and with the Genzel et al.~(1998) and Laurent et al.~(2000) mid-IR diagnostic diagrams.
Considering the whole sample, our results can be compared with the optical classification
and with L-band and hard X-ray studies, when available.
A substantial agreement is obtained in all cases. It is worth noticing that
the optical classification alone gives incomplete information:
all the sources classified as Seyferts show clear traces of AGN activity,
but 7 out of 8 among the ULIRGs with $\alpha_\mathit{bol}>0.25$ and
$\tau>1$ are indeed classified as LINERs or H~\textsc{ii} regions.
LINERs are again confirmed to be rather heterogeneous with respect to the
nature of their energy source. Such ambiguities can be solved by applying
our diagnostic.

\begin{figure}
\includegraphics[width=8.5cm]{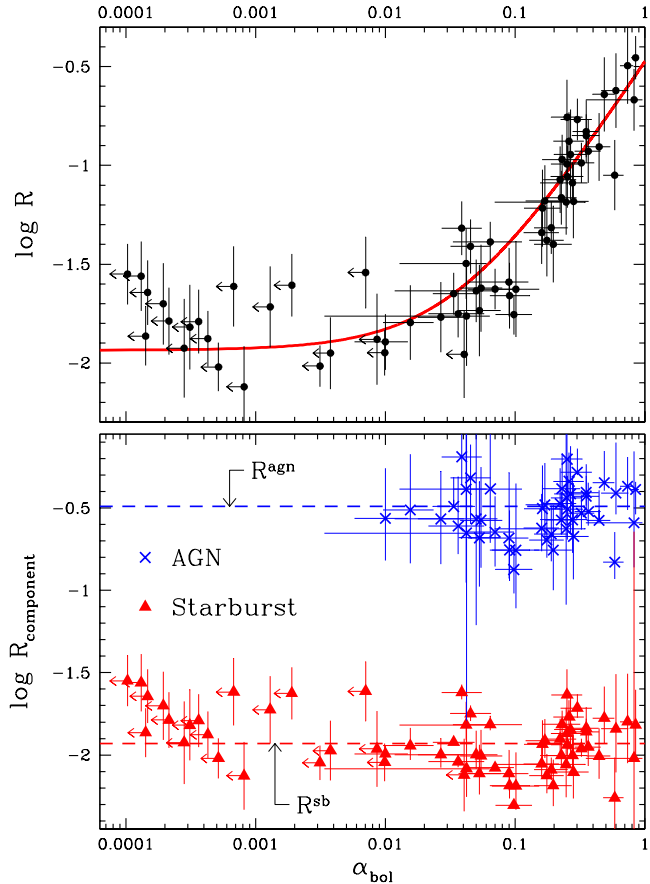}
\caption{\textit{(a)} Ratio $R$ between absorption-corrected
6-$\mu$m luminosity and bolometric luminosity,
versus the AGN bolometric contribution $\alpha_\mathit{bol}$. The error
bars of $R$ are due to the uncertainties in the total infrared flux $F_{IR}$
and in the intrinsic AGN fraction, $\alpha_6$.
The solid line is the best fit of the $R$--$\alpha_6$ relation
from Eq.\ref{e3} (plotted as a function of $\alpha_{\mathit{bol}}$ using Eq.\ref{e5}). \textit{(b)} Same
as above, with the AGN and SB components plotted separately.}
\label{f3}
\end{figure}

By inverting Eq.\ref{e5} the relation between $R$ and $\alpha_\mathit{bol}$ takes the neat form $R=\alpha_\mathit{bol}R^\mathit{agn}+(1-\alpha_\mathit{bol})R^\mathit{sb}$, and is plotted in Fig.\ref{f3}a.
As a final check we have computed for each source the following quantities:
\begin{equation}
\widehat R^\mathit{agn}=\left(\frac{\nu_6 \alpha_6 f_6^\mathit{int}}{\alpha_\mathit{bol} F_\mathit{IR}}\right)  \ \ \ \textrm{and} \ \ \
\widehat R^\mathit{sb}=\left[\frac{\nu_6 (1-\alpha_6) f_6^\mathit{int}}{(1-\alpha_\mathit{bol}) F_\mathit{IR}}\right].
\end{equation}

The results are shown in Fig.\ref{f3}b and
prove that our decomposition method is reliable in estimating
the AGN/SB contributions both to the 6-$\mu$m and to the
bolometric luminosity of local ULIRGs. In fact, the
estimated 6-$\mu$m to bolometric ratios of the SB component
in the composite sources (which are located at the bottom
right of the plot and are heavily dependent on our modeling)
are fully consistent, within the errors, with the
ratios of the pure starbursts (located at the bottom left and
directly computed from the measured 6-$\mu$m and IRAS fluxes).
This success is promising in anticipation of a forthcoming
study about the role of black hole accretion and star formation in the
intense infrared activity characterising the distant galaxies.

However, it is important to keep in mind the main limitations of our approach:\\
1) The narrow wavelength range used in this work simplifies
the decomposition analysis, but prevents us from a complete
study of the dust composition, density and geometrical distribution.
These elements strongly affect the overall mid-IR emission
longward of the silicate absorption feature, and can be investigated only
through an analysis of the whole IRS spectrum.\\
2) While the 5--8~$\mu$m templates seem to have little dispersion
(as discussed in detail in Section~\ref{3}), the spread in the 6-$\mu$m to
bolometric ratios $R^\mathit{agn}$ and $R^\mathit{sb}$ can be much higher, making
our estimates of the AGN/SB bolometric fractions more
uncertain than those in the 5--8~$\mu$m band. The uncertainties
in $\alpha_\mathit{bol}$ reported in Tab.\ref{t1} are obtained assuming
the mean ratios, with the errors on the mean given in Eq.\ref{e4}.
However, the dispersion around the best fit in Fig.\ref{f3}a
is significantly larger. We therefore consider this dispersion (0.3
dex, constant at all values of $\alpha_\mathit{bol}$) as the actual
uncertainty in the bolometric ratios of the individual sources.
The numerical results on the individuale sources are anyway precise enough to
estabilish which is the dominant source of the observed luminosity.

Overall, if we consider as a confidence limit the value
$\alpha_\mathit{bol}=0.01$ (i.e. the dispersion around our starburst
template), an AGN is present in 46 of the 68 ULIRGs
in our sample (including several of those optically classified
as H~\textsc{ii} regions), but it is significant ($\alpha_\mathit{bol}\ga0.25$)
only in $\sim$30\% of the cases. The SB process is responsible
for almost 90\% of the observed infrared luminosity of ULIRGs,
with no significant (i.e. $>5\%$) bias due to the sample selection.
A similar fraction holds for the subsample of 34 sources optically
classified as LINERs. Our analysis is also consistent with the
findings about the nature of high-redshift infrared-bright galaxies
detected in 24~$\mu$m \textit{Spitzer} MIPS surveys. IRS spectroscopy
shows that they are mostly $z\sim$1--3 galaxies, with an apparent bias
toward AGN-dominated sources (Houck et al. 2005). This is in agreement
with the $\sim$30 times higher AGN relative emission in the 5--8~$\mu$m
rest-frame wavelength range we have pointed out.

\section{Conclusions}
The use of average templates for AGN and SB emission
has allowed us to disentangle the two components in the 5--8~$\mu$m
spectra of 68 local ULIRGs, observed with the \textit{Spitzer Space Telescope}.
We have been able to detect an AGN in more than 60\% of our sources,
and estimate its contribution to the bolometric luminosity. In a
statistical sense, we confirm that local Ultraluminous Infrared Galaxies
are powered for $\sim$85\% by intense star formation and
for the remaining $\sim$15\% by AGN activity.
Our method proves to be successful in unveiling an intrinsically faint
or obscured AGN inside a ULIRG. In this context we also put on a sound
basis our initial assumption that the wavelength interval 5--8~$\mu$m
is an appropriate spectral range in order to search for AGNs:
an AGN turns out to be approximately 30 times more luminous at
6~$\mu$m than a starburst with the same bolometric luminosity.
\section*{Acknowledgments}
We are grateful to the anonymous referee for his/her
helpful and constructive comments.
We acknowledge financial support from the PRIN-MIUR 2006025203 grant
and the ASI-INAF grant I/023/05/0.


\label{lastpage}

\end{document}